\documentclass[conference,letterpaper]{IEEEtran}
\IEEEoverridecommandlockouts

\usepackage[T1]{fontenc}
\usepackage{cite}
\usepackage{amsmath,amssymb,amsfonts}
\usepackage{graphicx}
\usepackage{textcomp}
\usepackage{xcolor}
\usepackage{algorithm}
\usepackage{algpseudocode}
\usepackage{booktabs}
\usepackage{array}
\usepackage[hyphens]{url}
\usepackage[hidelinks]{hyperref}
\usepackage{fontawesome5}
\usepackage{hyperref}
\usepackage{eso-pic}
\AddToShipoutPictureBG*{%
  \AtPageLowerLeft{%
    \raisebox{0.35in}{%
      \makebox[\paperwidth]{%
        \parbox{0.85\paperwidth}{\centering\scriptsize
          \copyright~2026 IEEE. Personal use of this material is permitted.
          Permission from IEEE must be obtained for all other uses, in any
          current or future media, including reprinting/republishing this
          material for advertising or promotional purposes, creating new
          collective works, for resale or redistribution to servers or lists,
          or reuse of any copyrighted component of this work in other works.}}}}}

\def\BibTeX{{\rm B\kern-.05em{\sc i\kern-.025em b}\kern-.08em
    T\kern-.1667em\lower.7ex\hbox{E}\kern-.125emX}}

\let\OldIEEEauthorblockN\IEEEauthorblockN
\let\OldIEEEauthorblockA\IEEEauthorblockA

\renewcommand{\IEEEauthorblockN}[1]{%
  \OldIEEEauthorblockN{\fontsize{9}{10}\selectfont #1}%
}

\renewcommand{\IEEEauthorblockA}[1]{%
  \OldIEEEauthorblockA{\fontsize{6}{7}\selectfont #1}%
}

\begin{document}

\title{Stabilizing Autoregressive PDE Foundation Model Rollouts with Event-Triggered Context Healing}

\author{\IEEEauthorblockN{1\textsuperscript{st} Jaewan Park}
\IEEEauthorblockA{\textit{Mechanical Science and Engineering} \\
\textit{University of Illinois Urbana-Champaign}\\
Urbana, USA \\
jaewanp2@illinois.edu}
\and
\IEEEauthorblockN{2\textsuperscript{nd} Jay Phil Yoo}
\IEEEauthorblockA{\textit{Nuclear, Plasma \& Radiological Engineering} \\
\textit{University of Illinois Urbana-Champaign}\\
Urbana, USA \\
jayyoo2@illinois.edu}
\and
\IEEEauthorblockN{3\textsuperscript{rd} Kazuma Kobayashi}
\IEEEauthorblockA{\textit{Nuclear, Plasma \& Radiological Engineering} \\
\textit{University of Illinois Urbana-Champaign}\\
Urbana, USA \\
kazumak2@illinois.edu}
\and
\IEEEauthorblockN{4\textsuperscript{th} Seid Koric}
\IEEEauthorblockA{\textit{National Center for Supercomputing Applications} \\
\textit{University of Illinois Urbana-Champaign}\\
Urbana, USA \\
koric@illinois.edu}
\and
\IEEEauthorblockN{5\textsuperscript{th} Syed Bahauddin Alam}
\IEEEauthorblockA{\textit{Nuclear, Plasma \& Radiological Engineering} \\
\textit{University of Illinois Urbana-Champaign}\\
Urbana, USA \\
alams@illinois.edu}
\and
\IEEEauthorblockN{6\textsuperscript{th} Iwona Jasiuk}
\IEEEauthorblockA{\textit{Mechanical Science and Engineering} \\
\textit{University of Illinois Urbana-Champaign}\\
Urbana, USA \\
ijasiuk@illinois.edu}

\and
\IEEEauthorblockN{7\textsuperscript{th} Souvik Chakraborty}
\IEEEauthorblockA{\textit{Yardi School of Artificial Intelligence} \\
\textit{Indian Institute of Technology Delhi}\\
New Delhi, India \\
souvik@am.iitd.ac.in}

\and
\IEEEauthorblockN{8\textsuperscript{th} Diab W. Abueidda}
\IEEEauthorblockA{\textit{National Center for Supercomputing Applications} \\
\textit{University of Illinois Urbana-Champaign}\\
Urbana, USA \\
abueidd2@illinois.edu}

}

\maketitle

\begin{abstract}
Autoregressive PDE foundation models enable fast full-field forecasting but accumulate errors as predicted states are recursively reused. We introduce Event-Triggered Context Healing (ETCH), which compares forecasts with measurements from 12 cylinder-surface pressure taps. When their discrepancy exceeds a threshold, pressure-conditioned hourglass diffusion transformers reconstruct the velocity--pressure context for subsequent predictions. On 43 RealPDEBench Cylinder trajectories, ETCH reduces velocity and pressure errors from 48.05 and 63.71\% to 6.29 and 7.25\% for DPOT, and from 74.97 and 76.91\% to 5.16 and 6.84\% for Poseidon-T, using the same reconstructors and threshold. For DPOT, correcting 14.56\% of windows maintains errors close to reconstructing every window and supports average real-time forecasting and correction on an A100 GPU. These results show that sparse physical feedback stabilizes long autoregressive rollouts across PDE foundation models while reducing reconstruction cost. 

\end{abstract}


\vspace{0.8em}
\noindent\textbf{Code Availability:} Available at \href{https://github.com/benjamin0303/ETCH-Stabilizing-Autoregressive-PDE-Foundation-Model-Rollouts-with-Event-Triggered-Context-Healing}{\faGithub~ETCH Repository}.

\section{Introduction}

High-fidelity computational fluid dynamics (CFD) solvers provide reliable solutions to governing partial differential equations, but repeatedly running them for many initial conditions, parameter settings, or design queries can be prohibitively expensive. Learned PDE surrogates reduce this cost through fast neural-network inference; however, conventional surrogates are typically trained for a single equation, resolution, and parameter regime. PDE foundation models (PDE-FM) amortize this cost by pretraining on diverse PDE datasets to learn reusable representations that can be adapted to downstream systems with limited task-specific data. Models such as Walrus, DPOT, and Poseidon~\cite{Mccabe2025walrus,hao2024dpot,herde2024poseidon} follow this paradigm by transferring pretrained representations to downstream forecasting tasks.

Despite these advantages, fast inference and transferable pretraining do not guarantee stable long-horizon prediction. Many PDE foundation models generate a future block from a finite history and then roll forward autoregressively using that prediction as the context for the next model call. The model is therefore progressively conditioned on its own imperfect outputs. Errors introduced in one block become part of the next input, causing initially accurate trajectories to drift over long rollouts.

\begin{figure*}[t]
  \centering
  \includegraphics[width=\textwidth]{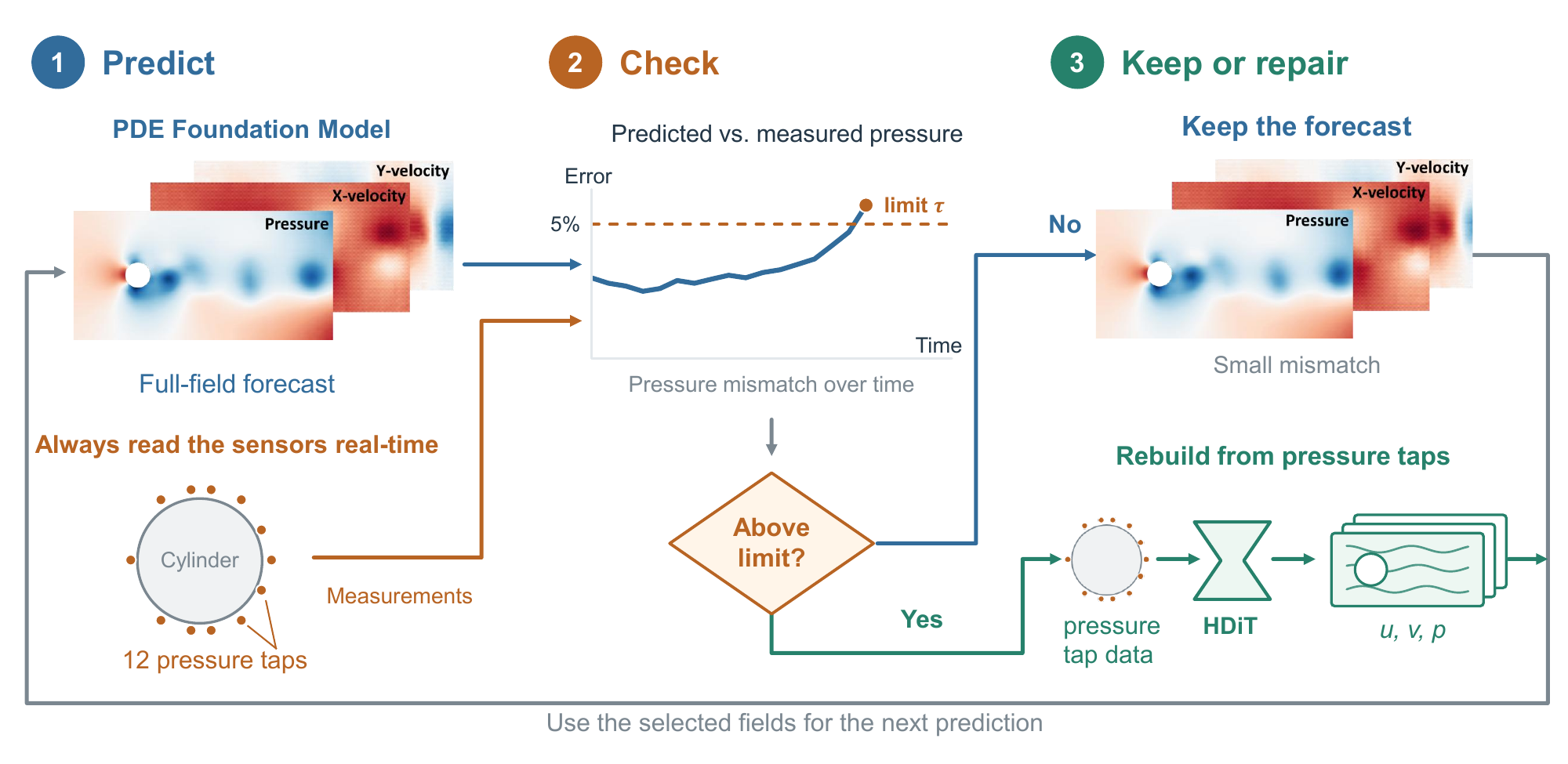}
  \caption{ETCH in three steps. Predict the full field, check the forecast
  against 12 pressure taps, and reconstruct only when their mismatch exceeds
  $\tau=0.05$. Otherwise, keep the forecast. HDiT uses tap histories and the
  known Reynolds number to reconstruct the latest 20 observed frames, without
  the current forecast as an input. The selected fields update DPOT's 20-frame
  context or Poseidon's single-frame anchor. The check follows arrival of the
  current observation window. Field stacks and the HDiT icon are schematic;
  the mismatch curve illustrates the trigger, with an orange dot marking
  a correction. The tap positions follow the evaluated sensor layout.}
  \label{fig:overview}
\end{figure*}

\begin{figure*}[t]
  \centering
  \includegraphics[width=\textwidth]{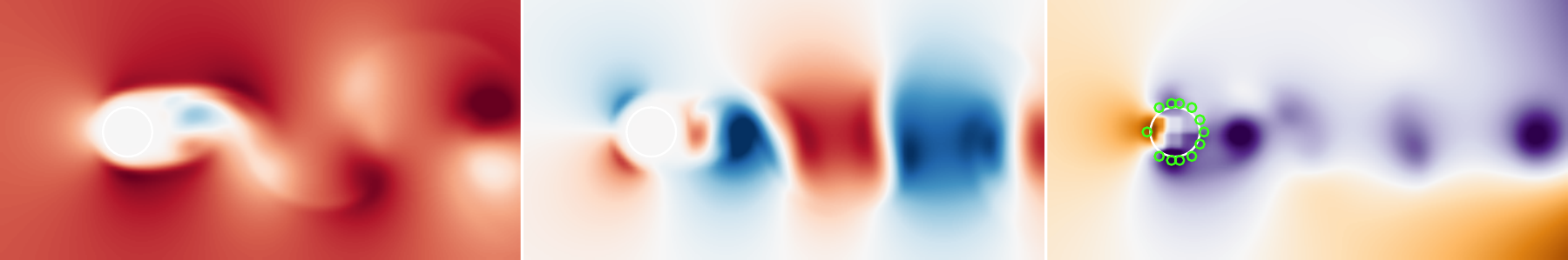}
  \caption{Representative RealPDEBench Cylinder fields: $x$-velocity, $y$-velocity, and pressure, from left to right. White circles mark the cylinder; pressure-tap locations are shown on the pressure field.}
  \label{fig:cylinder_fields}
\end{figure*}

Correcting drift requires information independent of the model-generated
trajectory, but continuous full-field measurement is rarely feasible.
Wall-pressure taps provide a practical observation interface for flow around
a solid body. Online correction must recover velocity and pressure throughout
the domain from these sparse boundary measurements while keeping pace with
the evolving flow.

Generative models can recover this missing state by learning spatial and
cross-field relationships conditioned on sparse measurements and physical
parameters~\cite{rozet2023sda,shysheya2024conditional,huang2024diffda, elata2025invfusion}.
Diffusion models support observation conditioning and measurement consistency,
but full-field reconstruction costs more than PDE-FM forecasting. Reconstructing
every window can therefore create a growing computational backlog.

These considerations motivate Event-Triggered Context Healing (ETCH;
Fig.~\ref{fig:overview}). The PDE-FM advances autoregressively while 12 cylinder
taps provide pressure every $5\,\mathrm{ms}$ ($200\,\mathrm{Hz}$).
After each observation window, a monitor compares measured and predicted
pressures. When their discrepancy exceeds a threshold, a generative model
reconstructs the rolling velocity--pressure context, updating both the state
estimate and the next forecast input. Every measurement contributes to
monitoring, while only selected cycles invoke full-state reconstruction.
We evaluate ETCH on unsteady cylinder flow using DPOT and
Poseidon~\cite{hao2024dpot,herde2024poseidon} with conditional Hourglass
Diffusion Transformer (HDiT) reconstructors~\cite{crowson2024hdit, elata2025invfusion}.
We compare reconstruction accuracy, long-horizon drift, and computation time
against uncorrected rollouts with HDiT or exact-CFD initialization and against
every-window reconstruction. The results show that selective reconstruction
stabilizes both forecasters, with DPOT meeting the average real-time compute
budget on an A100 GPU.

Our main contributions are threefold:
\begin{itemize}
\item We formulate PDE foundation model stabilization as a rolling-window sparse-observation feedback problem driven by pressure measurements acquired at high frequency.

\item We introduce ETCH, which uses a low-cost forecast-observation discrepancy to identify when the PDE foundation model state requires re-anchoring and invokes a full-state reconstructor only for the selected forecast cycles.

\item We develop a staged prior-removal curriculum that introduces temporal pressure-tap and Reynolds-number conditioning while withdrawing the forecast prior, yielding dual-branch HDiT reconstructors that recover full-field velocity and pressure without prior conditioning at deployment.
\end{itemize}

\section{Related Work}

\subsection{PDE Foundation Models and Autoregressive Forecasting}

DPOT combines Fourier attention with autoregressive denoising
pretraining~\cite{hao2024dpot}. Poseidon learns solution operators through a
multiscale transformer and semigroup-based training~\cite{herde2024poseidon},
while Walrus transfers across continuum-dynamics scenarios and improves
long-horizon stability~\cite{Mccabe2025walrus}. These methods improve transfer
and open-loop prediction through architecture or training. ETCH complements
them by using sparse physical measurements to detect drift and selectively
re-anchor a frozen forecaster during deployment.

\subsection{Sparse-Observation and Event-Triggered Estimation}

Variational and sequential data assimilation correct model predictions using
incomplete observations~\cite{ledimet1986variational,evensen1994sequential}.
Learning-based approaches recover fields through inverse mappings, reduced-order
representations, sensor placement, shallow decoders, or spatial encodings of
measurements~\cite{frerix2021variational,williams2023sensing,
everson1995gappy,manohar2018sensor,erichson2020shallow,fukami2021voronoi}.

Our setting additionally requires cross-field inference from boundary pressure
to full-field velocity and pressure. Event-based estimators use uncertainty or
measurement innovation to schedule updates or transmissions~\cite{trimpe2014event}.
ETCH retains continuous sensing and communication; the forecast-observation
discrepancy controls only whether full-field reconstruction is computed.

\subsection{Generative Models for Full-Field Reconstruction}

Score-based Data Assimilation and CoCoGen generate states conditioned on
incomplete observations or PDE information~\cite{rozet2023sda,jacobsen2023cocogen}.
Conditional diffusion supports forecasting and sparse
assimilation~\cite{shysheya2024conditional}; InvFusion incorporates the
degradation operator into a trained denoiser~\cite{elata2025invfusion}; and
DiffDA couples forecasts with sparse atmospheric observations in recurring
assimilation cycles~\cite{huang2024diffda}.

ETCH selectively invokes generative reconstruction to control its cost at
high observation rates. ANCHOR and divergence-aware CFD correction also
monitor surrogate reliability but recall numerical solvers for
correction~\cite{roy2025anchor,divergence2026cfd}. ETCH instead reconstructs
the corrective state directly from pressure taps online, without requiring full-field observation.

\section{Methods}
\label{sec:methods}

\subsection{Event-Triggered Blockwise Rollout and State Re-Anchoring}
\label{sec:closed_loop}

Our framework couples a frozen PDE foundation model with a sparse-observation
reconstructor in a blockwise closed loop. DPOT, a PDE-FM,  forecasts the full flow state, whereas HDiT, a generative reconstructor, initializes and selectively re-anchors the rollout using pressure
measurements. Let
\begin{equation}
X_t=[u_t,v_t,p_t]\in\mathbb{R}^{H\times W\times3}
\end{equation}
denote the full state and
\begin{equation}
y_t
=
H_p(X_t)+\varepsilon_t
\in\mathbb{R}^{m}
\label{eq:observation}
\end{equation}
denote the available observation, where $H_p$ extracts pressure at $m$ fixed
sensor locations and $\varepsilon_t$ represents measurement noise. The Reynolds
number $Re$ is assumed known.

The measurements are partitioned into non-overlapping blocks of $K$ frames.
The initial warm-up block is
\begin{equation}
\mathbf{Y}_0
=
[y_1,\ldots,y_K]
\in\mathbb{R}^{K\times m}.
\label{eq:warmup_observation_block}
\end{equation}
For each operational block $b=1,\ldots,B$, let $t_b=bK$ and define
\begin{align}
\mathbf{X}_b
&=
[X_{t_b+1},\ldots,X_{t_b+K}],
\\
\mathbf{Y}_b
&=
[y_{t_b+1},\ldots,y_{t_b+K}]
\in\mathbb{R}^{K\times m}.
\label{eq:observation_block}
\end{align}
Thus, $\mathbf{Y}_0$ provides the preceding sensor history and
$\mathbf{Y}_1$ corresponds to the first operational state block.

Before autoregressive forecasting begins, the composite HDiT reconstructor
$G_\phi$ uses the first two observation blocks and the known Reynolds number to
construct the initial full-state context:
\begin{equation}
\widetilde{\mathbf{X}}_1
=
\mathbf{X}_1^{\mathrm{rec}}
=
G_\phi
\left(
\mathbf{Y}_0,
\mathbf{Y}_1,
Re
\right).
\label{eq:initial_context}
\end{equation}
Given the selected context from the preceding block, the frozen DPOT forecaster
then predicts
\begin{equation}
\widehat{\mathbf{X}}_b
=
F
\left(
\widetilde{\mathbf{X}}_{b-1}
\right),
\qquad
b=2,\ldots,B.
\label{eq:FM_forecast}
\end{equation}

Once the corresponding observation block $\mathbf{Y}_b$ becomes available, the
predicted pressure is transformed to physical units and sampled at the sensor
locations. Let $H_p^{(K)}$ apply $H_p$ to all $K$ frames. We compute the
blockwise relative pressure discrepancy
\begin{equation}
s_b
=
\frac{
\left\|
H_p^{(K)}
\left(
\widehat{\mathbf{X}}_b
\right)
-
\mathbf{Y}_b
\right\|_2
}{
\max\!\left(
\left\|\mathbf{Y}_b\right\|_2,
\epsilon
\right)
},
\qquad
\epsilon=10^{-12},
\label{eq:block_trigger_score}
\end{equation}
where the norm spans all $K\times m$ pressure values. The correction decision is
\begin{equation}
a_b
=
\mathbb{I}\!\left[s_b>\tau\right],
\qquad
a_b\in\{0,1\}.
\label{eq:trigger_decision}
\end{equation}
The decision uses only the DPOT prediction and measured pressure taps; the
unobserved full state $\mathbf{X}_b$ is reserved for offline evaluation. We use
the fixed operating threshold $\tau=0.05$ in all reported experiments.

When $a_b=1$, HDiT reconstructs the current block from the preceding and current
pressure observations:
\begin{equation}
\mathbf{X}_b^{\mathrm{rec}}
=
G_\phi
\left(
\mathbf{Y}_{b-1},
\mathbf{Y}_b,
Re
\right)
\in
\mathbb{R}^{K\times H\times W\times3}.
\label{eq:composite_reconstruction}
\end{equation}
The operational state is then selected as
\begin{equation}
\widetilde{\mathbf{X}}_b
=
\begin{cases}
\widehat{\mathbf{X}}_b,
& a_b=0,
\\[2mm]
\mathbf{X}_b^{\mathrm{rec}},
& a_b=1.
\end{cases}
\label{eq:operational_block_update}
\end{equation}
Accordingly, an untriggered block retains the DPOT prediction, whereas a
triggered block is replaced by the HDiT reconstruction. The selected block
$\widetilde{\mathbf{X}}_b$ enters both the final assimilated trajectory and the
context for the next DPOT forecast.

Because the replacement is performed after the complete observation block has
arrived, a triggered block is a fixed-lag state estimate rather than an
ahead-of-time forecast. All reported closed-loop field errors are therefore
computed from the final sequence of selected blocks
$\{\widetilde{\mathbf{X}}_b\}$ after the triggered HDiT replacements.

\begin{table*}[t]
\centering
\caption{Prior-removal curricula for the velocity and pressure HDiT branches.}
\label{tab:hdit_curriculum}
\scriptsize
\begin{tabular}{@{}c c p{6.0cm} p{4.2cm} c@{}}
\toprule
Branch & Stage & Conditioning and prior transition
& Reconstruction target & Steps \\
\midrule
Velocity
& V1
& DPOT $[u,v,p]$, current pressure-tap, and $Re$
& $U_{\mathrm{CFD}}-U_{\mathrm{DPOT}}$
& 30k
\\
& V2
& DPOT $[u,v]$; DPOT $p\rightarrow\overline{p}$ during the first 10k steps;
current pressure taps and $Re$
& $U_{\mathrm{CFD}}-U_{\mathrm{DPOT}}$
& 20k
\\
& V3
& DPOT $[u,v]$, fixed $\overline{p}$, causal 20-frame tap history, and $Re$
& $U_{\mathrm{CFD}}-U_{\mathrm{DPOT}}$
& 20k
\\
& V4
& DPOT $[u,v]\rightarrow\overline{U}$ during the first 10k steps;
fixed $\overline{p}$, 20-frame taps, target offset, and $Re$
& $U_{\mathrm{CFD}}-U_{\mathrm{base}}$,
ending as $U_{\mathrm{CFD}}-\overline{U}$
& 30k
\\
\midrule
Pressure
& P1
& DPOT $[u,v,p]$, current pressure-tap, and $Re$
& $p_{\mathrm{CFD}}-p_{\mathrm{DPOT}}$
& 30k
\\
& P2
& DPOT $[u,v]$; DPOT $p\rightarrow\overline{p}$ during the first 5k steps;
current pressure taps and $Re$
& $p_{\mathrm{CFD}}-p_{\mathrm{base}}$,
ending as $p_{\mathrm{CFD}}-\overline{p}$
& 10k
\\
& P3
& DPOT $[u,v]$, fixed $\overline{p}$, causal 20-frame tap history, and $Re$
& $p_{\mathrm{CFD}}-\overline{p}$
& 10k
\\
& P4
& DPOT $[u,v]\rightarrow\overline{U}$ during the first 5k steps;
fixed $\overline{p}$, causal 20-frame tap history, and $Re$
& $p_{\mathrm{CFD}}-\overline{p}$
& 10k
\\
\bottomrule
\end{tabular}
\end{table*}

\subsection{Dual-Branch Full-State Reconstruction from Pressure Taps}
\label{sec:dual_hdit}

The composite reconstructor
$G_\phi=(G_{\phi_u},G_{\phi_p})$ consists of two independently trained
hourglass diffusion transformer (HDiT) branches. The velocity branch
$G_{\phi_u}$ reconstructs the two velocity components, whereas the pressure
branch $G_{\phi_p}$ reconstructs the pressure field. The branches use the same
backbone architecture but separate parameters and temporal conditioning because
velocity and pressure require different observation histories.

Let
\begin{equation}
U_t
=
[u_t,v_t]
\in
\mathbb{R}^{H\times W\times2}
\end{equation}
denote the velocity field. For the $j$th target frame in block $b$, we define the
target offset
\begin{equation}
\delta_j=j-1,
\qquad
j=1,\ldots,K.
\label{eq:target_offset}
\end{equation}
The velocity branch conditions every target on the complete pressure-tap block
$\mathbf{Y}_b$, the target offset, and the Reynolds number:
\begin{equation}
U_{t_b+j}^{\mathrm{rec}}
=
G_{\phi_u}
\left(
\mathbf{Y}_b,
\delta_j,
Re
\right)
\in
\mathbb{R}^{H\times W\times2}.
\label{eq:velocity_frame_reconstruction}
\end{equation}
Here, $\mathbf{Y}_b$ provides block-level temporal information, while
$\delta_j$ identifies the requested frame. All $K$ offsets are processed as one
inference batch to produce
\begin{equation}
\mathbf{U}_b^{\mathrm{rec}}
=
\left[
U_{t_b+1}^{\mathrm{rec}},
\ldots,
U_{t_b+K}^{\mathrm{rec}}
\right]
\in
\mathbb{R}^{K\times H\times W\times2}.
\label{eq:velocity_block_reconstruction}
\end{equation}
Because each target uses the complete observation block, the velocity branch
acts as a fixed-lag block smoother.

The pressure branch instead uses a causal $K$-frame pressure-tap history for
each target. For target index $j$, this history is
\begin{equation}
\mathbf{Y}_{b,j}^{\mathrm{hist}}
=
\left[
y_{t_b+j-K+1},
\ldots,
y_{t_b+j}
\right]
\in
\mathbb{R}^{K\times m}.
\label{eq:causal_tap_history}
\end{equation}
Histories for early frames draw from $\mathbf{Y}_{b-1}$, and no measurement
after the target frame is included. The pressure reconstruction is
\begin{equation}
p_{t_b+j}^{\mathrm{rec}}
=
G_{\phi_p}
\left(
\mathbf{Y}_{b,j}^{\mathrm{hist}},
Re
\right)
\in
\mathbb{R}^{H\times W}.
\label{eq:pressure_frame_reconstruction}
\end{equation}
The current observation $y_{t_b+j}$ is both the final element of the causal
history and spatially backprojected into the HDiT input. Although the histories
differ across target frames, all $K$ pressure queries are processed as one batch
after $\mathbf{Y}_b$ is complete:
\begin{equation}
\mathbf{P}_b^{\mathrm{rec}}
=
\left[
p_{t_b+1}^{\mathrm{rec}},
\ldots,
p_{t_b+K}^{\mathrm{rec}}
\right]
\in
\mathbb{R}^{K\times H\times W}.
\label{eq:pressure_block_reconstruction}
\end{equation}

For each query, the HDiT input contains nine channels: a three-channel noisy
state, a three-channel zero-filled spatial backprojection of the current
pressure taps, and a three-channel dense reference. A temporal adapter embeds
the corresponding pressure sequence, and a physical-parameter adapter encodes
the standardized Reynolds number. The velocity branch additionally includes
the target-offset embedding. The stochastic input and fixed dense reference are
suppressed in Eqs.~\eqref{eq:velocity_frame_reconstruction} and
\eqref{eq:pressure_frame_reconstruction} for clarity. Importantly, the dense
reference is shared across deployment samples and does not contain the current
DPOT prediction.

Finally, the two branch outputs are concatenated to form the reconstructed
full-state block:
\begin{equation}
\mathbf{X}_b^{\mathrm{rec}}
=
\operatorname{concat}
\left(
\mathbf{U}_b^{\mathrm{rec}},
\mathbf{P}_b^{\mathrm{rec}}
\right)
\in
\mathbb{R}^{K\times H\times W\times3}.
\label{eq:full_state_reconstruction}
\end{equation}
This separation enables sparse pressure histories to recover both interior
velocity and pressure without conditioning either deployed branch on the
potentially drifting DPOT state.

\subsection{Prior-Removal Training}
\label{sec:hdit_training}

The velocity and pressure reconstructors use independently parameterized copies of the same EDM-preconditioned HDiT backbone~\cite{karras2022edm, crowson2024hdit}. (Table~\ref{tab:hdit_curriculum}) Both branches are trained through progressive prior-removal curricula. During the initial stages, frozen DPOT predictions provide sample-specific spatial
references, allowing HDiT to first learn corrections around informative
full-field estimates. These predictions are then progressively replaced by
fixed reference fields computed from the training split, denoted by
$\overline{U}$ and $\overline{p}$. The final deployed branches therefore
reconstruct full fields from sparse pressure histories without receiving the
current DPOT prediction.

The velocity branch uses a mixed-noise EDM objective. The pressure branch is
fine-tuned with a pressure reconstruction loss and a tap-consistency term
that penalizes error at the observed locations. Both curricula first remove
the DPOT pressure reference, introduce temporal pressure histories, and then
remove the DPOT velocity reference. The final velocity stage also introduces
block-level conditioning and target offsets. We denote the final branches
by V4 and P4. Targets are residuals relative to their respective reference
fields. Each branch starts from an independent initialization, and weights
are transferred only between successive stages of that branch.

HDiT fields and residuals are encoded using statistics computed exclusively from
the training trajectories. For a physical field or residual $x$, we use
\begin{equation}
z_x
=
\frac{x-\mu_x}{4\sigma_x},
\qquad
x_{\mathrm{model}}
=
\frac{z_x+1}{2},
\label{eq:model_normalization}
\end{equation}
where $\mu_x$ and $\sigma_x$ are training-set statistics. The Reynolds number
uses $Re_{\mathrm{std}}=(Re-\mu_{Re})/\sigma_{Re}$.
The final pressure stages retain this affine transformation without clipping
the targets, tap histories, or reconstructed pressure values.


At deployment, the frozen branches reconstruct the target block from pressure
observations and $Re$ using EDM--Heun sampler; all $K$ target
queries of each branch are processed as one batch. Their predicted
physical-space corrections are added to the fixed references:
$U_{t_b+j}^{\mathrm{rec}}=\overline U+\Delta U_{t_b+j}$ and
$p_{t_b+j}^{\mathrm{rec}}=\overline p+\Delta p_{t_b+j}$.

\subsection{Cylinder Dataset and Implementation}
\label{sec:data}
\label{sec:implementation}

We use the numerical Cylinder subset of RealPDEBench \cite{hu2026realpdebench}
because its PIV data lack pressure. The CFD subset contains 92 trajectories
at different Reynolds numbers. For HDiT, we use a strictly trajectory-disjoint
split comprising 49 training trajectories and 43 held-out evaluation
trajectories. No frame from a held-out trajectory is used to train HDiT or
estimate its normalization and reference-field statistics. DPOT uses an existing
Cylinder-fine-tuned checkpoint whose associated normalizer includes all 92
trajectories, so its preprocessing is not strictly training-only.

Each held-out trajectory contains 3,990 full-state frames sampled at
$\Delta t=5\,\mathrm{ms}$. The original $128\times256$ velocity-pressure
fields are downsampled to $64\times128$. Each state
$X_t=[u_t,v_t,p_t]\in\mathbb{R}^{64\times128\times3}$ contains two velocity
components and gauge pressure. All pressure values retain the fixed reference
provided by the dataset. The observation operator $H_p$ extracts pressure at $m=12$ fixed,
wall-adjacent fluid cells surrounding the cylinder (Fig.~\ref{fig:cylinder_fields}). The same sensor layout is
used for every training and evaluation trajectory. Measurements are obtained
directly from the CFD pressure fields without additional synthetic noise;
therefore, $\varepsilon_t=0$ in Eq.~\eqref{eq:observation}. The trajectory
Reynolds number is treated as known and supplied to both HDiT branches.
The sensor rate is $200\,\mathrm{Hz}$. With $K=S=20$, the update interval is
$T_{\mathrm{update}}=S\Delta t=100\,\mathrm{ms}$.
For each held-out
trajectory, the first 20 frames are assigned to the preceding tap history
$\mathbf{Y}_0$, and the next 20 frames are assigned to $\mathbf{Y}_1$ for the initial
full-state reconstruction in Eq.~\eqref{eq:initial_context}.

We use the small DPOT configuration, denoted DPOT-S, as the PDE foundation
model. Its Cylinder-fine-tuned checkpoint uses
a 20-frame input context and a 20-frame prediction horizon. Closed-loop
evaluation advances this forecast by $S=20$ frames per cycle. Its parameters are frozen. Any DPOT predictions
used during the early HDiT curriculum stages are precomputed from this frozen
model, while the final V4 and P4 reconstructors operate without
sample-specific DPOT conditioning.

\subsection{Closed-Loop Evaluation and Runtime Protocol}
\label{sec:evaluation_protocol}

We compare three rollout policies for each foundation model on the same
held-out trajectories and pressure-tap layout. In \emph{HDiT initialization
only}, HDiT reconstructs $\widetilde{\mathbf{X}}_0$, after which the foundation
model proceeds without correction. \emph{Exact-CFD initialization} instead
uses the true initial context as an oracle-initialization reference.
\emph{Triggered correction} applies Eq.~\eqref{eq:operational_block_update}
with $\tau=0.05$. A shared \emph{HDiT-only} baseline reconstructs every window
directly from the pressure observations, without a foundation model.

Only complete $S$-frame observation windows are eligible for trigger evaluation and HDiT replacement. Frames in a final incomplete window remain DPOT predictions. Except for exact-CFD initialization and offline error computation, the held-out full-state fields are unavailable to the framework.

All errors are computed in physical units for velocity $U=[u,v]$, pressure $p$, and the full state $X=[u,v,p]$. Let $\mathcal{T}_{\mathrm{eval}}$ denote the frames following the initialization context. The primary accuracy metric is the mean per-frame relative error after the correction decision for each newly advanced interval:
\begin{equation}
E_q^{\mathrm{heal}}
=
\frac{1}{|\mathcal{T}_{\mathrm{eval}}|}
\sum_{t\in\mathcal{T}_{\mathrm{eval}}}
\frac{
\left\|
\widetilde{q}_t-q_t
\right\|_2
}{
\max(\|q_t\|_2,\epsilon)
},
\qquad
q\in\{U,p,X\}.
\label{eq:assimilated_state_error}
\end{equation}
Here, $\widetilde{q}_t$ is the DPOT output for an uncorrected cycle and the HDiT reconstruction for a corrected cycle. Thus, replaced DPOT fields are excluded from the reported closed-loop error. The initialization context is also excluded, whereas a final incomplete window is retained.

Standalone HDiT accuracy uses the same relative-error definition with
$q_t^{\mathrm{rec}}$ over the reconstructed-frame set $\mathcal{T}_{\mathrm{rec}}$.

Let $\mathcal{W}_{\mathrm{all}}$ contain all forecast windows and let $\mathcal{W}_{\mathrm{elig}}\subseteq\mathcal{W}_{\mathrm{all}}$ contain the complete windows eligible for correction. The number and fractions of applied corrections are
\begin{align}
N_{\mathrm{heal}} = \sum_{b\in\mathcal{W}_{\mathrm{elig}}}a_b,\qquad
\rho_{\mathrm{all}}=\frac{N_{\mathrm{heal}}}{|\mathcal{W}_{\mathrm{all}}|},
\qquad
\rho_{\mathrm{elig}} = 
\frac{N_{\mathrm{heal}}}{|\mathcal{W}_{\mathrm{elig}}|}.
\label{eq:correction_fractions}
\end{align}
These quantities measure HDiT usage relative to every-window reconstruction.
Runtime is measured on an NVIDIA A100 GPU with $40\,\mathrm{GB}$ of memory
after warm-up. For each complete cycle $b$, the service time $c_b$ includes
discrepancy evaluation, trigger selection, any HDiT reconstruction, context
update, and the next foundation-model forecast, including data transfer.
Sensor I/O, startup, and offline error computation are excluded. Each
trajectory contributes 197 complete cycles; its final partial interval is
scored for accuracy but excluded from timing. With
$T_{\mathrm{update}}=100\,\mathrm{ms}$, the recurring compute-to-stream ratio
(CSR) is
\begin{equation}
\mathrm{CSR}_{\mathrm{rec}}=
\frac{
\displaystyle\sum_{b=1}^{N_{\mathrm{cyc}}}c_b
}{
N_{\mathrm{cyc}} \cdot T_{\mathrm{update}}
},
\label{eq:recurring_csr}
\end{equation}
where $N_{\mathrm{cyc}}$ is the number of complete cycles. A value below one
means that average computation fits the update period. This supports
soft-real-time operation while permitting individual cycles to exceed
their deadlines.

\begin{figure*}[!t]
  \centering
  \includegraphics[width=\textwidth]{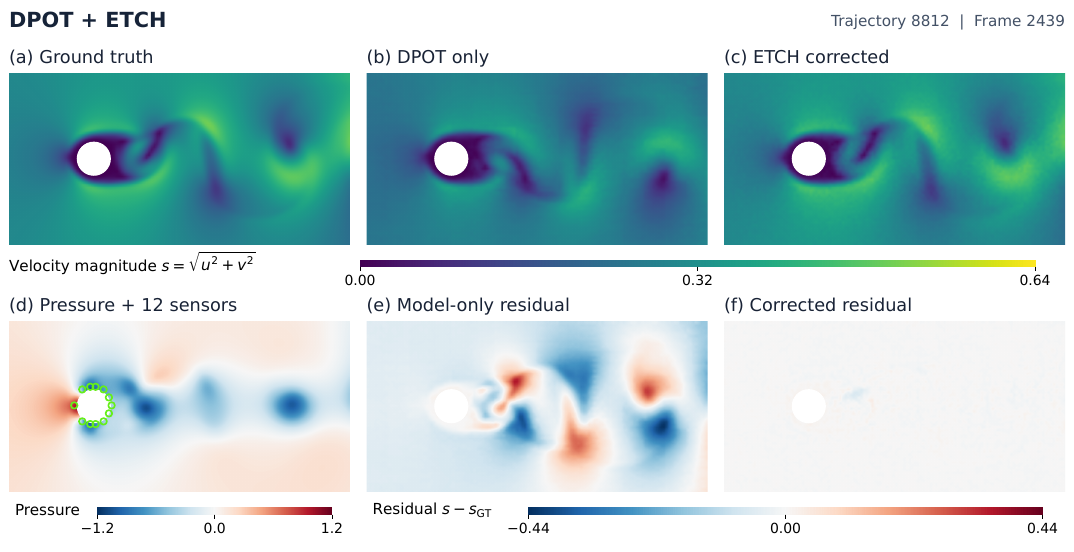}
  \caption{Comparison of DPOT rollouts with and without ETCH, starting from the same HDiT-initialized context. Velocity magnitude is computed from the velocity components as $s=\sqrt{u^2+v^2}$. (a) Ground truth; (b) DPOT prediction without online correction; (c) ETCH-corrected field; (d) ground-truth pressure with the 12 sensor locations. Panels (e,f) show the signed velocity-magnitude residuals, $s-s_{\mathrm{GT}}$, for (b,c), respectively. White circles indicate the cylinder.}
  \label{fig:dpot_comparison}
\end{figure*}

\section{Results}
\label{sec:results}

\subsection{Standalone Full-State Reconstruction from Pressure Taps}
\label{sec:results_reconstruction}

The deployed V4 and P4 branches achieve velocity and pressure errors of
$4.62\%$ and $5.93\%$, respectively, using only pressure observations, fixed
references, and $Re$ (Table~\ref{tab:closed_loop_results}). Both branches
reconstruct the full state without sample-specific forecast fields. This
provides a shared reconstruction reference for evaluating how much accuracy
ETCH retains when it invokes HDiT selectively.

\subsection{Closed-Loop Drift Suppression by Triggered Correction}
\label{sec:results_closed_loop}

Accurate initialization alone does not prevent long-horizon DPOT drift. With
HDiT initialization but no online correction, the velocity and pressure errors
increase to $48.05\%$ and $63.71\%$, respectively. Exact-CFD initialization
produces nearly identical degradation, confirming that the dominant failure
arises from autoregressive error accumulation rather than initialization error.

\begin{figure*}[!t]
  \centering
  \includegraphics[width=0.95\textwidth]{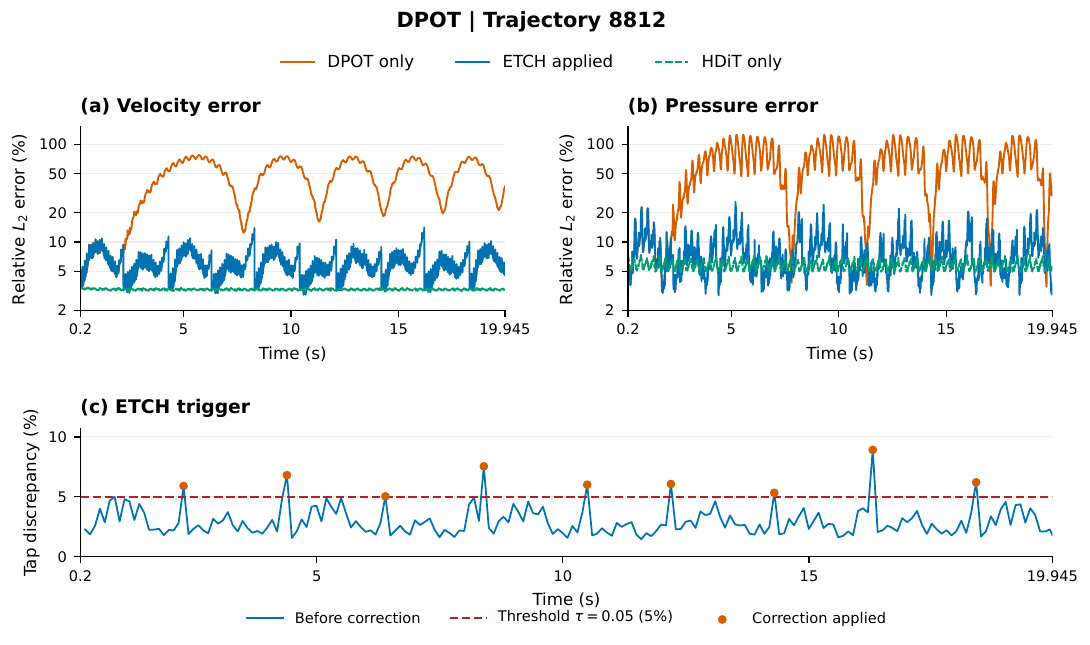}
  \caption{DPOT error history for Cylinder trajectory 8812 through the final
  frame ($19.945\,\mathrm{s}$). Panels (a,b) compare velocity $[u,v]$ and
  pressure relative $L_2$ errors for the HDiT-initialized uncorrected rollout,
  ETCH post-assimilation states, and HDiT-only reconstruction (logarithmic axes).
  Panel (c) shows pre-correction pressure-tap discrepancy at window endpoints:
  the dashed line marks $\tau=0.05$ (5\%); orange points mark corrections.}
  \label{fig:dpot_error_time}
\end{figure*}

With $\tau=0.05$, ETCH replaces $14.56\%$ of the eligible forecast windows and
reduces velocity and pressure errors to $6.29\%$ and $7.25\%$.
As shown in Table~\ref{tab:closed_loop_results}, these results remain
close to reconstructing every window while avoiding $85.44\%$ of the
corresponding online HDiT calls. Selective state replacement therefore retains
most of the stabilization benefit at substantially lower reconstruction cost.
Figure~\ref{fig:dpot_comparison} illustrates the accumulated drift in an
uncorrected DPOT rollout and the ETCH-corrected field at the same frame.
Figure~\ref{fig:dpot_error_time} follows trajectory 8812 through its final
stored frame, showing how selective corrections limit drift across the rollout.

\begin{table*}[!t]
\centering
\caption{Field accuracy on 43 evaluation trajectories with stride 20.
Relative $L_2$ errors are the mean $\pm$ standard deviation of trajectory
means over all 3,950 forecast frames through the final stored frame.
ETCH errors use fields after correction. Online HDiT rates exclude
initialization and use 8,471 complete windows per model.
HDiT-only uses the shared reconstructors for $[u,v,p]$ without
a foundation model.}
\label{tab:closed_loop_results}
\small
\begingroup
\renewcommand{\arraystretch}{1.15}
\begin{tabular}{@{}p{4.7cm} c c c c@{}}
\toprule
Configuration
& Online HDiT rate
& Velocity
& Pressure
& Full state
\tabularnewline
\midrule
HDiT-only reconstruction
& 100.00\%
& $4.62\pm2.04\%$
& $5.93\pm0.23\%$
& $5.18\pm0.60\%$
\tabularnewline

\midrule
\multicolumn{5}{@{}l}{\textit{DPOT-S}}\\
\midrule

\textbf{ETCH:} Triggered at $\tau=0.05$
& 14.56\%
& $6.29\pm1.91\%$
& $7.25\pm0.92\%$
& $6.58\pm0.75\%$
\tabularnewline

\midrule

HDiT initialization only
& 0\%
& $48.05\pm5.89\%$
& $63.71\pm4.91\%$
& $57.90\pm5.62\%$
\tabularnewline

Exact-CFD initialization only
& 0\%
& $48.64\pm6.22\%$
& $64.30\pm4.86\%$
& $58.47\pm5.56\%$
\tabularnewline

\midrule
\multicolumn{5}{@{}l}{\textit{Poseidon-T}}\\
\midrule

\textbf{ETCH:} Triggered at $\tau=0.05$
& 27.33\%
& $5.16\pm1.63\%$
& $6.84\pm1.21\%$
& $6.11\pm1.41\%$
\tabularnewline

HDiT initialization only
& 0\%
& $74.97\pm22.97\%$
& $76.91\pm12.77\%$
& $77.33\pm14.68\%$
\tabularnewline

Exact-CFD initialization only
& 0\%
& $75.12\pm22.91\%$
& $77.05\pm12.66\%$
& $77.48\pm14.57\%$
\tabularnewline

\bottomrule
\end{tabular}
\endgroup
\end{table*}

\subsection{Soft-Real-Time Performance}
\label{sec:results_runtime}

\begin{table}[t]
\centering
\caption{Recurring compute on an A100 with stride 20 and a
100 ms update period. Times average 8,471 complete windows;
CSR is mean time divided by the update period.
ETCH includes monitoring, optional HDiT reconstruction, context update,
and the next forecast, including transfers. HDiT-only reconstructs
$[u,v,p]$ without a foundation model. Sensor I/O, startup, warm-up,
and offline error calculation are excluded.}
\label{tab:runtime_summary}
\small
\begin{tabular}{@{}l r r@{}}
\toprule
Configuration & Mean (ms) & CSR\\
\midrule
DPOT-S + ETCH & 31.06 & 0.311\\
Poseidon-T + ETCH & 92.52 & 0.925\\
HDiT-only & 109.57 & 1.096\\
\bottomrule
\end{tabular}
\end{table}

At stride 20, DPOT with ETCH averages $31.06\,\mathrm{ms}$ per cycle,
giving $\mathrm{CSR}_{\mathrm{rec}}=0.311$
(Table~\ref{tab:runtime_summary}). This includes both forecasting and
the selected reconstructions. HDiT-only reconstruction averages
$109.57\,\mathrm{ms}$, giving $\mathrm{CSR}_{\mathrm{rec}}=1.096$;
reconstructing every window therefore accumulates computational delay
under the same $100\,\mathrm{ms}$ budget. ETCH preserves sufficient average
throughput by allocating reconstruction to the windows that need it.

The threshold $\tau$ and update period are application-dependent choices.
They can be adjusted together to balance reconstruction accuracy, correction
frequency, and the available real-time compute budget. The reported
$\tau=0.05$ and $100\,\mathrm{ms}$ period provide one common operating point
for comparing the two foundation models.

\subsection{Transfer to Poseidon-T}
\label{sec:poseidon}

\begin{figure*}[!t]
  \centering
  \includegraphics[width=\textwidth]{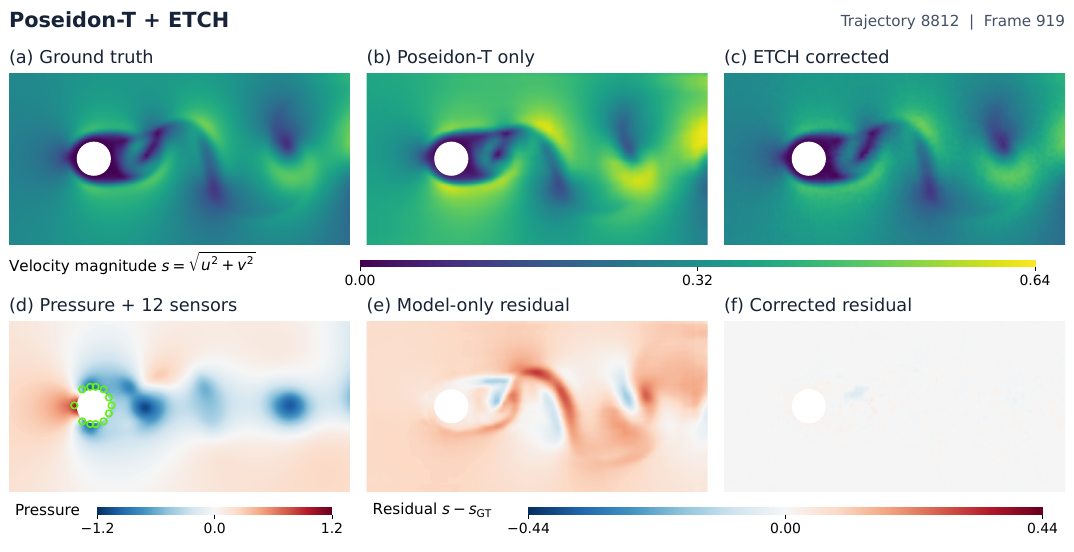}
  \caption{Comparison of Poseidon-T rollouts with and without ETCH. Panel ordering and field definitions follow the DPOT comparison in Fig.~\ref{fig:dpot_comparison}.}
  \label{fig:poseidon_comparison}
\end{figure*}

We replace DPOT with Poseidon-T~\cite{herde2024poseidon} while reusing the frozen
V4/P4 reconstructors and $\tau=0.05$. No HDiT retraining or threshold retuning
is needed because reconstruction depends on sensor inputs, not forecast fields.
Poseidon-T is fine-tuned for 5,000 updates on 49 training trajectories and
queries leads 1--20 from one anchor frame. Fields are
spectrally resized between $64\times128$ and $128\times128$. At stride $S$,
predicted lead $S$, or the last reconstructed frame, becomes the next anchor.
Evaluation covers all 3,950 forecast frames in each of 43 trajectories.

Table~\ref{tab:closed_loop_results} compares policies at stride 20.
With HDiT initialization, ETCH reduces velocity and pressure
errors from $74.97\%$ and $76.91\%$ without online correction to $5.16\%$
and $6.84\%$. Exact-CFD initialization alone also fails to prevent drift.
ETCH corrects $27.33\%$ of complete windows, avoiding $72.67\%$ of online
HDiT calls relative to reconstructing every window, whose errors are $4.62\%$
and $5.93\%$. Figure~\ref{fig:poseidon_comparison} shows the corresponding
field comparison. The shared reconstructors therefore stabilize both forecasters.
At this operating point, Poseidon-T with ETCH averages 92.52\,ms per cycle
($\mathrm{CSR}_{\mathrm{rec}}=0.925$), within the 100\,ms budget. Its cost
exceeds that of DPOT because Poseidon-T queries leads 1--20 from a single
anchor frame, and because it triggers reconstruction more often.

\section{Conclusion}
\label{sec:discussion_conclusion}

ETCH stabilizes long autoregressive DPOT and Poseidon-T rollouts through sparse
pressure feedback and selective full-state reconstruction. The same frozen
HDiT branches and threshold keep post-assimilation errors close to every-window
reconstruction while avoiding $85.44\%$ and $72.67\%$ of online HDiT calls,
respectively. Accurate initialization alone does not prevent either model's
drift. On an A100 GPU, both DPOT and Poseidon with ETCH fits the average $100\,\mathrm{ms}$ compute budget at stride 20. Threshold and update-period choices allow the accuracy--computation tradeoff to reflect the needs of the application.

The present evaluation is limited to noiseless, fixed-location pressure taps
in the RealPDEBench Cylinder family, two PDE foundation models, and a fixed
threshold that was not independently optimized. Future work should examine
measurement noise and missing sensors, validate threshold selection, and test
transfer across geometries, flow regimes, and PDE foundation models.

\section*{Acknowledgment}

This work is supported by the NSF Cybershuttle project (NSF Award No.\ 2209875). Also the Delta and DeltaAI (NSF awards OAC-2005572 and OAC-2320345 and the State of Illinois) resources were utilized for this research. Delta and DeltaAI are state-of-the art compute resources from joint efforts of the University of Illinois Urbana-Champaign and the National Center for Supercomputing Applications (NCSA). The authors greatly appreciate the help from the Research Consulting Directorate from NCSA, and the Center for Artificial Intelligence Innovation (CAII).

\bibliographystyle{IEEEtran}
\bibliography{references}

@inproceedings{herde2024poseidon,
  author    = {Herde, Maximilian and Raoni{\'c}, Bogdan and Rohner, Tobias
               and K{\"a}ppeli, Roger and Molinaro, Roberto
               and {de B{\'e}zenac}, Emmanuel and Mishra, Siddhartha},
  title     = {{Poseidon}: Efficient Foundation Models for {PDEs}},
  booktitle = {Advances in Neural Information Processing Systems},
  volume    = {37},
  year      = {2024},
  doi       = {10.52202/079017-2311}
}

@misc{Mccabe2025walrus,
  author = {McCabe, Michael and others},
  title  = {{Walrus}: A Cross-Domain Foundation Model for Continuum Dynamics},
  year   = {2025},
  note   = {arXiv:2511.15684}
}

@article{ledimet1986variational,
  author  = {{Le Dimet}, F.-X. and Talagrand, O.},
  title   = {Variational algorithms for analysis and assimilation of
             meteorological observations: theoretical aspects},
  journal = {Tellus A},
  volume  = {38},
  number  = {2},
  pages   = {97--110},
  year    = {1986},
  doi     = {10.1111/j.1600-0870.1986.tb00459.x}
}

@article{evensen1994sequential,
  author  = {Evensen, Geir},
  title   = {Sequential data assimilation with a nonlinear quasi-geostrophic
             model using {Monte Carlo} methods to forecast error statistics},
  journal = {Journal of Geophysical Research: Oceans},
  volume  = {99},
  number  = {C5},
  pages   = {10143--10162},
  year    = {1994},
  doi     = {10.1029/94JC00572}
}

@inproceedings{frerix2021variational,
  author    = {Frerix, Thomas and Kochkov, Dmitrii and Smith, Jamie
               and Cremers, Daniel and Brenner, Michael and Hoyer, Stephan},
  title     = {Variational Data Assimilation with a Learned Inverse
               Observation Operator},
  booktitle = {Proceedings of the 38th International Conference on Machine Learning},
  series    = {Proceedings of Machine Learning Research},
  volume    = {139},
  pages     = {3449--3458},
  year      = {2021},
  url       = {https://proceedings.mlr.press/v139/frerix21a.html}
}

@misc{williams2023sensing,
  author = {Williams, Jan P. and Zahn, Olivia and Kutz, J. Nathan},
  title  = {Sensing with shallow recurrent decoder networks},
  year   = {2023},
  note   = {arXiv:2301.12011}
}

@article{trimpe2014event,
  author  = {Trimpe, Sebastian and D'Andrea, Raffaello},
  title   = {Event-Based State Estimation With Variance-Based Triggering},
  journal = {IEEE Transactions on Automatic Control},
  volume  = {59},
  number  = {12},
  pages   = {3266--3281},
  year    = {2014},
  doi     = {10.1109/TAC.2014.2351951}
}

@article{jacobsen2023cocogen,
  author  = {Jacobsen, Christian and Zhuang, Yilin and Duraisamy, Karthik},
  title   = {{CoCoGen}: Physically Consistent and Conditioned Score-Based
             Generative Models for Forward and Inverse Problems},
  journal = {SIAM Journal on Scientific Computing},
  volume  = {47},
  number  = {2},
  pages   = {C399--C425},
  year    = {2025},
  doi     = {10.1137/24M1636071}
}

@inproceedings{elata2025invfusion,
  author    = {Elata, Noam and Chung, Hyungjin and Ye, Jong Chul
               and Michaeli, Tomer and Elad, Michael},
  title     = {{InvFusion}: Bridging Supervised and Zero-shot Diffusion
               for Inverse Problems},
  booktitle = {Advances in Neural Information Processing Systems},
  volume    = {38},
  year      = {2025},
  doi       = {10.52202/085713-1203}
}

@article{everson1995gappy,
  author  = {Everson, Richard and Sirovich, Lawrence},
  title   = {Karhunen--Lo\`{e}ve procedure for gappy data},
  journal = {Journal of the Optical Society of America A},
  volume  = {12},
  number  = {8},
  pages   = {1657--1664},
  year    = {1995}
}

@article{manohar2018sensor,
  author  = {Manohar, Krithika and Brunton, Bingni W. and
             Kutz, J. Nathan and Brunton, Steven L.},
  title   = {Data-driven sparse sensor placement for reconstruction:
             Demonstrating the benefits of exploiting known patterns},
  journal = {IEEE Control Systems Magazine},
  volume  = {38},
  number  = {3},
  pages   = {63--86},
  year    = {2018}
}

@article{erichson2020shallow,
  author  = {Erichson, N. Benjamin and Mathelin, Lionel and Yao, Zhewei
             and Brunton, Steven L. and Mahoney, Michael W. and
             Kutz, J. Nathan},
  title   = {Shallow neural networks for fluid flow reconstruction with
             limited sensors},
  journal = {Proceedings of the Royal Society A},
  volume  = {476},
  number  = {2238},
  pages   = {20200097},
  year    = {2020}
}

@article{fukami2021voronoi,
  author  = {Fukami, Kai and Maulik, Romit and Ramachandra, Nesar and
             Fukagata, Koji and Taira, Kunihiko},
  title   = {Global field reconstruction from sparse sensors with
             {V}oronoi tessellation-assisted deep learning},
  journal = {Nature Machine Intelligence},
  volume  = {3},
  number  = {11},
  pages   = {945--951},
  year    = {2021}
}

@inproceedings{hu2026realpdebench,
  author    = {Hu, Peiyan and Feng, Haodong and Liu, Hongyuan and
               Yan, Tongtong and Deng, Wenhao and Gao, Tianrun and
               Zheng, Rong and Zheng, Haoren and Yu, Chenglei and
               Wang, Chuanrui and Li, Kaiwen and Ma, Zhi-Ming and
               Zhou, Dezhi and Lu, Xingcai and Fan, Dixia and Wu, Tailin},
  title     = {{RealPDEBench}: A Benchmark for Complex Physical Systems
               with Real-World Data},
  booktitle = {International Conference on Learning Representations
               (ICLR)},
  year      = {2026},
  note      = {Oral Presentation}
}

@inproceedings{hao2024dpot,
  author    = {Hao, Zhongkai and others},
  title     = {{DPOT}: Auto-Regressive Denoising Operator Transformer for
               Large-Scale {PDE} Pre-Training},
  booktitle = {International Conference on Machine Learning (ICML)},
  year      = {2024}
}

@misc{roy2025anchor,
  author = {Roy, Rajyasri and Nayak, Dibyajyoti and Goswami, Somdatta},
  title  = {{ANCHOR}: Error-Controlled Adaptive Numerical Correction for
            Neural Operator Time Marching},
  note   = {arXiv:2512.19643v2},
  year   = {2026}
}

@misc{divergence2026cfd,
  author = {Zou, Xiangrui and Zhao, Zhuoqun and Barrag{\'a}n, Guillermo
            and {Le Clainche}, Soledad},
  title = {Divergence-aware adaptive prediction framework for
           accelerating {CFD} simulations of unsteady flows},
  note  = {arXiv:2605.24150},
  year  = {2026}
}

@inproceedings{rozet2023sda,
  author    = {Rozet, Fran\c{c}ois and Louppe, Gilles},
  title     = {Score-based Data Assimilation},
  booktitle = {Advances in Neural Information Processing Systems
               (NeurIPS)},
  year      = {2023},
  note      = {arXiv:2306.10574}
}

@inproceedings{shysheya2024conditional,
  author    = {Shysheya, Aliaksandra and Diaconu, Cristiana and
               Bergamin, Federico and Perdikaris, Paris and
               Hern\'{a}ndez-Lobato, Jos\'{e} Miguel and
               Turner, Richard E. and Mathieu, Emile},
  title     = {On conditional diffusion models for {PDE} simulations},
  booktitle = {Advances in Neural Information Processing Systems
               (NeurIPS)},
  year      = {2024},
  note      = {arXiv:2410.16415}
}

@inproceedings{huang2024diffda,
  author    = {Huang, Langwen and Gianinazzi, Lukas and Yu, Yuejiang and
               Dueben, Peter D. and Hoefler, Torsten},
  title     = {{DiffDA}: A Diffusion Model for Weather-scale Data
               Assimilation},
  booktitle = {Proceedings of the 41st International Conference on
               Machine Learning (ICML)},
  series    = {PMLR},
  volume    = {235},
  pages     = {19798--19815},
  year      = {2024},
  note      = {arXiv:2401.05932}
}

@inproceedings{crowson2024hdit,
  title     = {Scalable High-Resolution Pixel-Space Image Synthesis with Hourglass Diffusion Transformers},
  author    = {Crowson, Katherine and Baumann, Stefan Andreas and Birch, Alex and Abraham, Tanishq Mathew and Kaplan, Daniel Z. and Shippole, Enrico},
  booktitle = {Proceedings of the 41st International Conference on Machine Learning (ICML)},
  series    = {Proceedings of Machine Learning Research},
  volume    = {235},
  pages     = {9550--9575},
  year      = {2024},
  note      = {arXiv:2401.11605}
}

@inproceedings{karras2022edm,
  title     = {Elucidating the Design Space of Diffusion-Based Generative Models},
  author    = {Karras, Tero and Aittala, Miika and Aila, Timo and Laine, Samuli},
  booktitle = {Advances in Neural Information Processing Systems (NeurIPS)},
  volume    = {35},
  pages     = {26565--26577},
  year      = {2022},
  note      = {arXiv:2206.00364}
}

\end{document}